\documentclass[preprint,10pt]{elsarticle}

\usepackage{amssymb}
\usepackage{amsfonts}
\usepackage{bbm}
\usepackage{mathrsfs}
\usepackage{mathrsfs}
\usepackage{amsthm}
\usepackage{amsfonts,mathrsfs,latexsym,amsmath,amssymb,amsthm}

\journal{Discrete Mathematics}

\begin{document}

\begin{frontmatter}


 \title{Generalized Quasi-Cyclic Codes Over $\mathbb{F}_q+u\mathbb{F}_q$\tnoteref{label1}}

\title{Generalized Quasi-Cyclic Codes Over $\mathbb{F}_{q}+u \mathbb{F}_{q}$}


\author{Jian Gao,~Linzhi Shen,~Fang-Wei Fu }

\address{Chern Institute of Mathematics and LPMC, Nankai University, P. R. China}

\begin{abstract}
Generalized quasi-cyclic (GQC) codes with arbitrary lengths over the ring $\mathbb{F}_{q}+u\mathbb{F}_{q}$, where $u^2=0$, $q=p^n$, $n$ a positive integer and $p$ a prime number, are investigated. By the Chinese Remainder Theorem, structural properties and the decomposition of GQC codes are given. For $1$-generator GQC codes, minimal generating sets and lower bounds on the minimum distance are given. As a special class of GQC codes, quasi-cyclic (QC) codes over $\mathbb{F}_q+u\mathbb{F}_q$ are also discussed briefly in this paper.

\end{abstract}


\end{frontmatter}


\section{Introduction}
\vskip 3mm \noindent
Codes over finite rings have been studied since the early 1970s. There are a lot of works on codes over finite rings after the discovery that certain good nonlinear binary codes can be constructed from cyclic codes over $\mathbb{Z}_4$ via the Gary map \cite{Hammons}. Quasi-cyclic (QC) codes over finite rings constitute a remarkable generalization of cyclic codes \cite{Aydin,Bhaintwal,Cui,Ling 2,Siap 2}. More recently, they have produced many codes over finite fields which meet the best possible values of minimum distances of codes with the same lengths and dimensions \cite{Aydin, Bhaintwal, Siap 2}. The notion of generalized quasi-cyclic (GQC) codes over finite fields was introduced by Siap and Kulhan \cite{Siap 1}. Some further structural properties of such codes were studied by Esmaeili and Yari \cite{Esmaeili}. Based on structural properties of GQC codes, Esmaeili and Yari gave two construction methods (Construction A and Construction B) and some optimal and suboptimal GQC codes were obtained by these methods \cite{Esmaeili}. In \cite{Cao 1}, Cao studied the GQC codes of arbitrary length over finite fields. He investigated the structural properties of GQC codes and gave the explicit enumeration of all $1$-generator GQC codes and $1$-generator GQC codes with a fixed parity-check polynomial respectively. As a natural generalization, GQC codes over Galois rings were introduced by Cao, structural properties and explicit enumeration of GQC codes were also studied \cite{Cao 2}.

\vskip 3mm \noindent {\bf Definition 1.1} \cite[Definition 1.1]{Cao 2} \emph{Let $R$ be a commutative ring with identity and $m_1$, $m_2$, $\ldots$, $m_\ell$ be positive integers. Denote $R_i=R[x]/(x^{m_i}-1)$ for $i=1,2,\ldots,\ell$. Any $R[x]$-submodule of the $R[x]$-module ${\mathcal R}=R_1\times R_2\times \cdots \times R_\ell$ is called a generalized quasi-cyclic (GQC) code over $R$ of block length $(m_1, m_2, \ldots, m_\ell)$ and length $\sum_{i=1}^\ell m_i$.}

\vskip 3mm  If $\mathscr{C}$ is a GQC code over $R$ of block length $(m_1, m_2, \ldots, m_\ell)$ and length $\sum_{i=1}^\ell m_i$ with $m=m_1=m_2=\cdots=m_\ell$, then $\mathscr{C}$ is a \emph{quasi-cyclic} (QC) \emph{code} of length $m\ell$ over $R$. Furthermore, if $\ell=1$, then $\mathscr{C}$ is a \emph{cyclic code} of length $m$ over $R$.

\vskip 3mm For the QC code of length $m\ell$ with index $\ell$ over the finite chain ring $R$ satisfying ${\rm gcd}(m, \kappa)=1$, where $\kappa$ denotes the characteristic of $R$, Ling and Sol\'{e} \cite{Ling 2} decomposed the QC code by the Chinese Remainder Theorem (CRT) into product of shorter codes over some extensive rings of $R$. Bhaintwal and Wasan \cite{Bhaintwal} studied the QC code over the prime integer residue ring $\mathbbm{Z}_q$. They viewed a QC code of length $m\ell$ with index $\ell$ as an $\mathbb{Z}_q[x]/(x^m-1)$-submodule of ${\rm GR}(q,\ell)[x]/(x^m-1)$, where ${\rm GR}(q,\ell)$ was the $\ell$-th Galois extension ring of $\mathbb{Z}_q$. A sufficient condition for $1$-generator QC code to be $\mathbb{Z}_q$-free was given and some distance bounds for $1$-generator QC code were also discussed. In \cite{Cui}, Cui and Pei studied the $1$-generator quasi-cyclic code over $\mathbb{Z}_4$. Under some conditions, they gave the enumeration of quaternary $1$-generator QC code of length $m\ell$ with index $\ell$ and described an algorithm to obtain one and only one generator for each $1$-generator QC code. For the QC code of length $m\ell$ with index $\ell$ over finite chain ring $R=\mathbb{F}_2+u\mathbb{F}_2$, where $m$ is an arbitrary length, Siap et al. \cite{Siap 2} determined the type of $1$-generator QC code over $R$ and the size by giving a minimal generating set. They also determined the rank and introduced a lower bound for minimum Hamming distance of the free $1$-generator QC code over $R$.

\vskip 3mm For GQC codes of block length $(m_1, m_2, \ldots, m_\ell)$ and length $\sum_{i=1}^\ell m_i$ over Galois ring $R$ satisfying ${\rm gcd}(m_1, \kappa)={\rm gcd}(m_2, \kappa)=\cdots={\rm gcd}(m_\ell, \kappa)=1$, where $\kappa$ denotes the characteristic of $R$, Cao \cite{Cao 2} used CRT to give an $R[x]$-module isomorphism of the GQC code over $R$, and this led to an explicit enumeration formula of the GQC code. But structural properties of GQC codes of \underline{arbitrary lengths} over the ring $R=\mathbb{F}_q+u\mathbb{F}_q$ has not been considered to the best of our knowledge.

\vskip 3mm The main aim of this present paper is to study GQC codes of arbitrary lengths over $\mathbb{F}_q+u\mathbb{F}_q$, focusing on structural properties. In Section 2, some useful results which will be used in the following sections are presented. In Section 3, we investigate the structural properties of GQC codes of arbitrary lengths, giving the decomposition of GQC codes over $\mathbb{F}_q+u\mathbb{F}_q$. In Section 4, we mainly study $1$-generator GQC codes by giving the minimum generating sets and a lower bound on the minimum Lee (Hamming) distance of the free $1$-generator GQC codes. Using Gray map, we obtain some good (optimal or suboptimal) linear codes over finite fields. In Section 5, we discuss a special class of GQC codes called QC codes over $R$. We use another point of view to research QC codes. The duals of QC codes are also discussed briefly.

\section{Preliminaries}
\vskip 6mm \noindent
A finite commutative ring with identity is called a \emph{finite chain ring} if its ideals are linearly ordered by inclusion. It is well known that every ideal of finite chain ring is principal and its maximal ideal is unique. Let $R$ denote the finite chain ring and $\gamma$ a generator of its maximal ideal. The ideals of $R$ form a chain as follows
$$ ( 0 )=( \gamma^s)\subseteq( \gamma^{s-1})\subseteq \cdots \subseteq ( \gamma ) \subseteq ( 1 )=R.$$ The integer $s$ is called the \emph{nilpotency index} of $R$. If $R/(\gamma) \cong \mathbb{F}_{q}$, then $|R|=q^s$. For the class of finite commutative chain rings, we have the following equivalent conditions.

\vskip 3mm \noindent {\bf Proposition 2.1} \cite[Proposition 2.1]{Dinh} \emph{For a finite commutative ring $R$, the following conditions are equivalent: \\
$(i)$ ~$R$ is a local ring and the maximal ideal $M$ of $R$ is principal; \\
$(ii)$ ~$R$ is a local principal ideal ring; \\
$(iii)$ ~$R$ is a chain ring.}

\vskip 3mm The classical examples of finite chain rings which are not finite fields are the integer residue ring $\mathbb{Z}_{p^s}$, the Galois ring ${\rm GR}(p^m, s)$ and the ring $\mathbb{F}_{p^n}+u\mathbb{F}_{p^n}+\cdots+u^{s-1}\mathbb{F}_{p^n}$, where $u^s=0$, $p$ is a prime number and $n$, $s$ are positive integers such that $s\geq 2$. Note that the ring $R=\mathbb{F}_{p^n}+u\mathbb{F}_{p^n}+\cdots+u^{s-1}\mathbb{F}_{p^n}$ is isomorphic to $\mathbb{F}_{p^n}[u]/(u^s)$, the only finite chain ring with characteristic $p$ and nilpotency index $s$, and its maximal ideal is $(u)$. Define the ring epimorphism $^- : R\rightarrow R/(u)$ by $r\mapsto \overline{r}$, where $\overline{r}$ denotes $r+(u)$. Clearly, $R/(u)$ is the residue field $\mathbb{F}_{p^n}$. Extend the ring epimorphism $^- : R[x]\rightarrow (R/(u))[x]$ by $r_0+r_1x+\cdots+r_nx^n\mapsto \overline{r}_0+\overline{r}_1x+\cdots+\overline{r}_nx^n$. Then we denote the image of $f(x)\in R[x]$ under the map $^-$ as $\overline{f}(x)\in \overline{R}[x]$.

\vskip 3mm Let $f(x)$ and $g(x)$ be polynomials of $R[x]$. A monic polynomial $d(x)$ is called a \emph{greatest common divisor} of $f(x)$ and $g(x)$ if $d(x)$ is a divisor of $f(x)$ and $g(x)$; and if $e(x)$ is a divisor of $f(x)$ and $g(x)$, then $e(x)$ is a divisor of $d(x)$. We denote $d(x)={\rm gcd}(f(x), g(x))$. Two polynomials $f(x)$ and $g(x)$ are said to be \emph{coprime} over $R$ if there are two polynomials $a(x)$ and $b(x)$ in $R[x]$ such that $a(x)f(x)+b(x)g(x)=1$. It is to be noted that in $R[x]$ two coprime polynomials may have a common divisor with degree $\geq 1$. However, it is clearly that the common divisor must be a unit in $R[x]$. Therefore, if let $f(x)$ and $g(x)$ be the monic polynomials, then their common divisor is only $1$ if they are coprime of them. A polynomial $f(x)\in R[x]$ is said to be \emph{basic irreducible} (or \emph{basic primitive}) if $\overline{f}(x)$ is irreducible (or primitive) in $\overline{R}[x]$.

\vskip 3mm In this paper, we mainly consider GQC codes over the ring $R=\mathbb{F}_{q}+u\mathbb{F}_{q}$, where $u^2=0$, $q=p^n$, $p$ is a prime number and $n$ is a positive integer. In the rest of this section, we list some useful results which will be used in this paper.

\vskip 3mm \noindent {\bf Theorem 2.2 } \cite[Theorem 2.9]{Wan}\emph{ Let $f_1$, $f_2$, $\ldots$, $f_r$ be pairwise coprime monic polynomials of degree $\geqslant 1$ over $R$, $f=f_1f_2\ldots f_r$ and ${\mathcal R}_f=R[x]/(f)$. Let $\widehat{f}_i=f/f_i$. Then there exist $a_i$, $b_i\in R[x]$ such that $a_if_i+b_i\widehat{f}_i=1$. Let $e_i=b_i\widehat{f}_i+(f)$. Then \\
$(i)$~$e_1$, $e_2$, $\ldots$, $e_r$ are mutually orthogonal non-zero idempotents of ${\mathcal R}_f$;\\
$(ii)$~$1=e_1+e_2+\cdots+e_r$ in  ${\mathcal R}_f$;\\
$(iii)$~Let ${\mathcal R}_fe_i=(e_i)$ be the principal ideal of ${\mathcal R}_f$ generated by $e_i$. Then $e_i$ is the identity of  ${\mathcal R}_fe_i$ and  ${\mathcal R}_fe_i=(\widehat{f}_i+(f))$; \\
$(iv)$~${\mathcal R}_f=\oplus_{i=1}^r{\mathcal R}_fe_i$;\\
$(v)$~The map $R[x]/(f_i)\rightarrow {\mathcal R}_fe_i$ defined by $g+(f_i)\mapsto (g+(f))e_i$ is a well-defined isomorphism of rings;\\
$(vi)$~${\mathcal R}_f=R[x]/(f)\cong \bigoplus_{i=1}^rR[x]/(f_i)$.}

\vskip 3mm \noindent {\bf Theorem 2.3 } \cite[Theorem 1]{Abualrub} \emph{Let $\mathscr{C}$ be a cyclic code in $R[x]/(x^m-1)$ and $R=\mathbb{F}_q+u\mathbb{F}_q$, where $u^2=0$, $q=p^n$, $p$ is a prime number and $n$ is a positive integer. Then \\
$(1)$~ If ${\rm gcd}(p,m)=1$ then $R[x]/(x^m-1)$ is a principal ideal ring and $\mathscr{C}=(g(x), ua(x))=(g(x)+ua(x))$, where $g(x), a(x) \in \mathbb{F}_q[x]$ with $a(x)\mid g(x) \mid (x^m-1)$;\\
$(2)$~ If ${\rm gcd}(p,m)\neq1$ then \\
$(i)$~ $\mathscr{C}=(g(x)+up(x))$ where $g(x)\mid (x^m-1)$ over $\mathbb{F}_q$ and $(g(x)+up(x))\mid (x^m-1)$ over $R$, and $g(x)\mid p(x)(\frac{x^m-1}{g(x)})$; or \\
$(ii)$~ $\mathscr{C}=(g(x)+up(x), ua(x))$ where $g(x)$, $a(x)$ and $p(x)$ are polynomials with $a(x)\mid g(x) \mid (x^m-1)$, $a(x)\mid p(x)(\frac{x^m-1}{g(x)})$ and ${\rm deg }a(x)> {\rm deg} p(x)$. }

\section{Structural properties of GQC codes}
\vskip 6mm \noindent
Let $R=\mathbb{F}_q+u\mathbb{F}_q$, where $u^2=0$ and $q=p^n$, $p$ is a prime number and $n$ is a positive integer. Let $m_i=p^{e_i}\widetilde{m}_i$, where ${\rm gcd}(p, \widetilde{m}_i)=1$ for each $i=1,2,\ldots,\ell$. Then $x^{\widetilde{m}_i}-1$ has a unique factorization $x^{\widetilde{m}_i}-1=f_{i,1}f_{i,2}\ldots f_{i,s_i}$, where $f_{i,1}, f_{i,2}, \ldots, f_{i, s_i}$ are pairwise coprime monic basic irreducible polynomials over $R$, which implies $x^{m_i}-1=f_{i,1}^{p^{e_i}}f_{i,2}^{p^{e_i}}\ldots f_{i,s_i}^{p^{e_i}}$. Let $\{ g_1, g_2, \ldots, g_s\}=\{ f_{i,j} \mid 1\leq i\leq \ell, 1\leq j \leq s_i\}$. Then we have $$x^{m_i}-1=g_1^{d_{i,1}}g_2^{d_{i,2}}\ldots g_s^{d_{i,s}}$$ where $d_{i,k}=p^{e_i}$ if $g_k=f_{i,j}$ for some $1\leq j \leq s_i$, and $d_{i, k}=0$ if ${\rm gcd}(g_k, x^{m_i}-1)=1$, for all $1\leq i \leq \ell$ and $1\leq k \leq s$.

\vskip 3mm Suppose the set $\{ r_{k,1}, r_{k,2}, \ldots, r_{k, t_k}\}=\{ d_{i,k}\mid d_{i,k}\neq 0,  1 \leq i \leq \ell\}$, where $r_{k,1}>r_{k,2}>\cdots > r_{k,t_k}\geq {\rm min}\{ p^{e_i}\mid i=1,2, \ldots, \ell\}$ and $n_{k,t_k}=\mid \{ i\mid d_{i,k}=r_{k, t_k}\}\mid$ for $k=1,2,\ldots,\ell$. It is obviously that $n_{k,1}r_{k,1}+\cdots +n_{k,t_k}r_{k,t_k}=d_{1,k}+\cdots +d_{\ell, k}$. Let ${\mathcal M}_k=(R[x]/(g_k^{r_{k,1}}))^{n_{k,1}} \times \cdots \times (R[x]/(g_k^{r_{k,t_k}}))^{n_{k,t_k}}$. It is clear that $${\mathcal M_k}=\bigoplus_{d_{i,k}\neq0, 1\leq i\leq l}R[x]/(g_k^{d_{i,k}})=\bigoplus_{i=1}^\ell R[x]/(g_k^{d_{i,k}})$$ is up to an $R[x]$-module isomorphism.

\vskip 3mm \noindent {\bf Theorem 3.1 } \emph{Let ${\mathcal R}=R_1\times R_2\times\cdots \times R_\ell$, where $R_i=R[x]/(x^{m_i}-1)$ for all $i=1,2,\ldots,\ell$. Then there exists an $R[x]$-module isomorphism $\phi$ from ${\mathcal R}$ onto ${\mathcal M}_1\times {\mathcal M}_2 \times \cdots \times {\mathcal M}_s$ such that $\mathscr{C}$ is a GQC code of block length $(m_1, m_2, \ldots, m_\ell)$ and length $\sum_{i=1}^\ell m_i$ over $R$ if and only if for each $1\leq k \leq s $ there is an unique $R[x]$-module $M_k$ of ${\mathcal M}_k$ such that $\phi (\mathscr{C})=M_1\times M_2\times\cdots \times M_s$.}
\vskip 3mm \noindent\emph{ Proof} It is easy to check that ${\rm max}\{ d_{1,k}, d_{2,k}, \ldots, d_{\ell, k}\}=r_{k,1}$ for each $k=1,2,\ldots,s$. Denote $$g=g_1^{r_{1,1}}g_2^{r_{2,1}}\cdots g_s^{r_{s,1}},~ \widehat{g}_k=\frac{g}{g_k^{r_{k,1}}},$$ $$\widetilde{g}_{i,k}=\frac{x^{m_i}-1}{g_k^{d_{i,k}}}, ~i=1,2,\ldots,\ell,~k=1,2,\ldots,s.$$ Then there exist polynomials $u_{i,k}, w_{i,k}\in R[x]$ such that $$g_k^{r_{k,1}}=u_{i,k}g_k^{d_{i,k}}~{\rm and}~ \widehat{g}_k=w_{i,k}\widetilde{g}_{i,k},~i=1,2,\ldots,\ell,~k=1,2,\ldots,s.$$ Since $g_k^{r_{k,1}}$ and $\widehat{g}_k$ are coprime, there exist polynomials $b_k,~c_k\in R[x]$ such that $b_k\widehat{g}_k+c_kg_k^{r_{k,1}}=1$, which implies that $b_kw_{i,k}\widetilde{g}_{i,k}+c_ku_{i,k}g_k^{d_{i,k}}=1$ in $R[x]$. Let $\varepsilon_{i,k}=b_kw_{i,k}\widetilde{g}_{i,k}+(x^{m_i}-1)=b_k\widehat{g}_k+(x^{m_i}-1)\in R_i$. Then from Theorem 2.2, we have

\vskip 3mm

(i)~$\varepsilon_{i,k}=0$ if and only if ${\rm gcd}(g_k, x^{m_i}-1)=1,~k=1,2,\ldots,s.$

\label{}

(ii)~$\varepsilon_{i,1},\varepsilon_{i,2},\ldots,\varepsilon_{i,s}$ are mutually orthogonal idempotents of $R_i$.

\label{}

(iii)~$\varepsilon_{i,1}+\varepsilon_{i,2}+\cdots+\varepsilon_{i,s}=1$ in $R_i$.

\label{}

(iv)~Let $R_{i,k}=R_i\varepsilon_{i,k}$ be the principle ideal of $R_i$ generated by $\varepsilon_{i,k}$. Then $\varepsilon_{i,k}$ is the identity of $R_{i,k}$ and $R_{i,k}=b_k\widehat{g}_kR_i$. Hence $R_{i,k}=\{ 0\}$ if and only if ${\rm gcd}(g_k, x^{m_i}-1)=1$.

\label{}

(v)~$R_i=\oplus_{k=1}^sR_{i,j}$.

\label{}

(vi)~For each $k=1,2,\ldots,s$, the mapping $\phi_{i,k}:~R_{i,k}\rightarrow R[x]/(g_k^{d_{i,k}})$, defined by $$\phi_{i,k}:~fb_k\widehat{g}_k+(x^{m_i}-1)\mapsto f+(g_k^{d_{i,k}}), f\in R[x]$$ is a well defined isomorphism of rings.

\label{}

(vii)~$R_i=R[x]/(x^{m_i}-1)\simeq \bigoplus_{j=1}^sR[x]/(g_j^{d_{i,j}})$.

\vskip 3mm From (vi), we have a well defined $R[x]$-module isomorphism $\Phi_k$ from $b_k\widehat{g}_k{\mathcal R}$ onto $R[x]/(g_k^{d_{1,k}})\times \cdots \times R[x]/(g_k^{d_{\ell, k}})$, which defined by $$\Phi_k:~(\alpha_1, \ldots, \alpha_\ell)\mapsto (\phi_{1,k}(\alpha_1),\ldots,\phi_{\ell, k}(\alpha_\ell)), \alpha_i\in R_{ik}, i=1,2,\ldots,\ell.$$ $\Phi_k$ can introduce a natural $R[x]$-module isomorphism $\mu_k$ from $b_k\widehat{g}_k{\mathcal R}$ onto ${\mathcal M}_k$.

\vskip 3mm For any $c=(c_0,c_1,\ldots,c_\ell)\in {\mathcal R}$, from (v) we deduce $c=(b_1\widehat{g}_1c_1+\cdots+b_s\widehat{g}_sc_1, \ldots, \\ b_1\widehat{g}_1c_\ell+\cdots+b_s\widehat{g}_sc_\ell)=b_1\widehat{g}_1c+\cdots+b_s\widehat{g}_sc$, where $b_k\widehat{g}_kc\in b_k\widehat{g}_kR_1\times\cdots\times b_k\widehat{g}_kR_\ell$ for all $k=1,2,\ldots,s$. Hence ${\mathcal R}=b_1\widehat{g}_1{\mathcal R}+\cdots+b_s\widehat{g}_s{\mathcal R}$. Let $c_1$, $c_2$, $\ldots$, $c_s\in {\mathcal R}$ satisfying $b_1\widehat{g}_1c_1+\cdots+b_s\widehat{g}_sc_s=0$. By $(x^{m_i}-1)\mid g$ for all $i=1,2,\ldots,\ell$, it follows that $g{\mathcal R}=\{0\}$. Then for each $k=1,2,\ldots,s$, from $b_k\widehat{g}_k+c_kg_k^{r_{k,1}}=1$, $g=g_k^{r_{k,1}}\widehat{g}_k$ and $g\mid \widehat{g}_\tau \widehat{g}\sigma$ for all $1\leq \tau\neq \sigma \leq s$, we deduce $b_k\widehat{g}_kc_k=0$. Hence ${\mathcal R}=\bigoplus_{j=1}^sb_j\widehat{g}_j{\mathcal R}$.

\vskip 3mm Define $\phi:~\beta_1+\beta_2+\cdots+\beta_s\mapsto (\mu_1(\beta_1), \mu_2(\beta_2), \ldots, \mu_s(\beta_s)), \beta_k\in b_k\widehat{g}_k{\mathcal R}, k=1,2,\ldots,s$. Then $\phi$ is an $R[x]$-module isomorphism from ${\mathcal R}$ onto ${\mathcal M}_1\times\cdots\times{\mathcal M}_s$. For any $R[x]$-module $M_j$, it is obvious that $M_1\times\cdots\times M_s$ is an $R[x]$-module of ${\mathcal M}_1\times\cdots\times{\mathcal M}_s$. Therefore there is a unique $R[x]$-submodule $\mathscr{C}$ of ${\mathcal R}$ such that $\phi(\mathscr{C})=M_1\times\cdots\times M_s$.      \hfill $\Box$

\vskip 3mm Since ${\mathcal M_k}=\bigoplus_{d_{i,k}\neq0, 1\leq i\leq \ell}R[x]/(g_k^{d_{i,k}})=\bigoplus_{i=1}^\ell R[x]/(g_k^{d_{i,k}})$ is up to an $R[x]$-module isomorphism, Theorem 3.1 can lead to a decomposition of the GQC code as follows.

\vskip 3mm \noindent {\bf Corollary 3.2 } \emph{Let $\mathscr{C}$ be a GQC code of block length $(m_1, m_2, \ldots, m_\ell)$ and length $\sum_{i=1}^\ell m_i$ over $R$. Then
$$\mathscr{C}=\bigoplus_{i=1}^s\mathscr{C}_i,$$ where $\mathscr{C}_i$, $1\leq i\leq s$, is an $R[x]$-submodule of $R[x]/(g_i^{d_1,i})\times \cdots \times R[x]/(g_i^{d_{\ell, i}})$ and each $j$-th, $1\leq j\leq \ell$, component in $\mathscr{C}_i$ is zero if $d_{j,i}=0$ or an element of the ring $R[x]/(g_i^{d_{j,i}})$.}        \hfill $\Box$

\vskip 3mm A GQC code $\mathscr{C}$ of block length $(m_1, m_2, \ldots, m_\ell)$ and length $\sum_{i=1}^\ell m_i$ is called \emph{$\rho$-generator} over $R$ if $\rho$ is the smallest positive integer for which there are codewords $A_i(x)=(a_{i,1}(x), a_{i,2}(x), \ldots, a_{i,\ell}(x))$, $1\leq i \leq \rho$, in $\mathscr{C}$ such that $\mathscr{C}=R[x]A_1(x)+R[x]A_2(x)+\cdots +R[x]A_\rho(x)$.

\vskip 3mm Assume that $m_j=p^an_j$, where ${\rm gcd}(p, n_j)=1$, for all $j=1,2,\ldots,\ell$ and each $\mathscr{C}_i$ is free, $i=1,2,\ldots,s$, with rank $k_i$. Let ${\mathcal K}={\rm max }\{ k_i \mid 1\leq i\leq s\}$.

\vskip 3mm \noindent {\bf Theorem 3.3 } \emph{Let $\mathscr{C}$ be a $\rho$-generator GQC code of block length $(m_1, m_2, \ldots, m_\ell)$ and length $\sum_{i=1}^\ell m_i$ over $R$. Let $m_j=p^an_j$, where ${\rm gcd}(p, n_j)=1$, for all $j=1,2,\ldots,\ell$ and $\mathscr{C}=\bigoplus_{i=1}^s\mathscr{C}_i$, where each $\mathscr{C}_i$ be free, $i=1,2,\ldots,s$, with rank $k_i$. Then $\rho={\mathcal K}$. In fact, any GQC code $\mathscr{C}$ with $\mathscr{C}=\bigoplus_{i=1}^s\mathscr{C}_i$, where each $\mathscr{C}_i$ is free, $i=1,2,\ldots,s$, with rank $k_i$ satisfying $\rho={\rm max} k_i$, is a $\rho$-generator GQC code.}

\vskip 3mm \noindent\emph{ Proof} Let $\mathscr{C}$ be a $\rho$-generator GQC code generated by the elements $A_j(x)=(a_{j,1}(x), a_{j,2}(x), \ldots, a_{j,l}(x)) \in {\mathcal R}, ~j=1,2,\ldots,\rho$. Then for each $i=1,2,\ldots,s$, $\mathscr{C}_i$ is spanned as an $R[x]$-module by $\widetilde{A}_{(j)}(x)=(\widetilde{a}_{j,1}(x), \widetilde{a}_{j,2}(x), \ldots, \widetilde{a}_{j,\ell}(x))$, where $\widetilde{a}_{j,\nu}(x)=a_{j,\nu}(x)~({\rm mod}\; g_i^{p^a})$ if $g_i^{p^a}$ is a factor of $x^{m_i}-1$ or $\widetilde{a}_{j,\nu}(x)=0$ otherwise, $\nu=1,2,\ldots,\ell$. Hence $k_i\leq \rho$ for each $i$, and so ${\mathcal K}\leq \rho$.
\vskip 1mm On the other hand, since ${\mathcal K}={\rm max}k_i$, there exist $Q_{i,j}(x)\in R[x]^\ell$, $1\leq j \leq {\mathcal K}$, such that $Q_{i,j}(x)$ span $\mathscr{C}_i$, $1\leq i \leq s$, as an $R[x]$-module. Then from Corollary 3.2, for each $1\leq j \leq {\mathcal K}$, there exists ${\mathcal Q}_{j}(x)\in \mathscr{C}$ such that ${\mathcal Q}_j(x)=Q_{i,j}(x)~({\rm mod}\; g_i^{p^a})$ and $\mathscr{C}$ is generated by ${\mathcal Q}_j(x)$, $1\leq j \leq{\mathcal K}$. Hence $\rho \leq {\mathcal K}$, which implies that $\rho={\mathcal K}$.                                    \hfill $\Box$

\vskip 3mm Now let $m_j=p^an_j$, where ${\rm gcd}(p, n_j)=1$ and $\mathscr{C}=\bigoplus_{i=1}^s\mathscr{C}_i$, where each $\mathscr{C}_i$ is free, $i=1,2,\ldots,s$, with rank $k_i$. If $\mathscr{C}$ is a $1$-generator GQC code of block length $(m_1, m_2, \ldots, m_\ell)$ and length $\sum_{i=1}^\ell m_i$ over $R$, then from Theorem 3.3, each $\mathscr{C}_i$, $i=1,2,\ldots,s$, is either trivial or an $[\ell, 1]$ linear code over $R[x]/(g_i^{p^a})$. Conversely, any GQC code $\mathscr{C}$ with each $\mathscr{C}_i$ is free with rank at most $1$ is a $1$-generator GQC code.

\vskip 3mm \noindent {\bf Example 3.4 } Let $R=\mathbb{F}_3+u\mathbb{F}_3$ and ${\mathcal R}=R[x]/(x^6-1)\times R[x]/(x^{12}-1)$. Consider the $2$-generator GQC code of block length $(6,12)$ and length $6+12=18$ generated by $A_1(x)=(x^4-1, x^2-x)$ and $A_2(x)=(x^3, x^2+1)$ over $R$. Since $x^6-1=(x-1)^3(x+1)^3$ and $x^{12}-1=(x-1)^3(x+1)^3(x^2+1)^3$ over $R$, from Theorem 3.1, $${\mathcal R}\cong (R[x]/((x-1)^3))^2\times (R[x]/((x+1)^3))^2\times R[x]/((x^2+1)^3).$$ Then up to an $R[x]$-module isomorphism
 \begin{equation}
  \begin{array}{ccc}
    {\mathcal R} & \cong & (R[x]/((x-1)^3), R[x]/((x-1)^3))\\
     & \bigoplus & (R[x]/((x+1)^3), R[x]/((x+1)^3))\\
     & \bigoplus & (0, R[x]/((x^2+1)^3)).\\
  \end{array}
\end{equation}
This implies that the GQC code $\mathscr{C}$ can be decomposed into $\mathscr{C}=\bigoplus_{i=1}^3\mathscr{C}_i$, where $\mathscr{C}_1$ is the $[2,2]$ linear code over $R[x]/((x-1)^3)$, $\mathscr{C}_2$ is the $[2,2]$ linear code over $R[x]/((x+1)^3)$ and $\mathscr{C}_3$ is the $[2,1]$ linear code over $R[x]/((x^2+1)^3)$. Let $k_i$ be the rank of $\mathscr{C}_i$, $i=1,2,3$. Then $${\rm max}k_i=2={\rm the ~number~of~generators~of~}\mathscr{C}.$$

\section{$1$-generator GQC codes}
\vskip 6mm \noindent
Let $\mathscr{C}$ be a $1$-generator GQC code of block length $(m_1, m_2, \ldots, m_\ell)$ and length $\sum_{i=1}^\ell m_i$ with generator $(f_1, f_2, \ldots, f_\ell)$, where $f_i\in R_i, ~i=1,2,\ldots,\ell$, over $R$. Define a well defined $R[x]$-homomorphism $\varphi_i$ from ${\mathcal R}$ onto $R_i$ such that $\varphi_i(f_1, f_2, \ldots, f_\ell)=f_i$. Then $\varphi_i(\mathscr{C})$ is a cyclic code of $R_i$. From Theorem 2.3, we have
\vskip 1mm
$(i)$~If ${\rm gcd}(p, m_i)=1$, then $f_i\in \varphi_i(\mathscr{C})$ can be selected to be of the form $f_i=b_i(x)(g_i(x)+ua_i(x))$, where $b_i(x)\in R[x]$ and $g_i(x), a_i(x)\in \mathbb{F}_q[x]$ with $a_i(x)\mid g_i(x)\mid (x^{m_i}-1)$;
\vskip 1mm
$(ii)$~If ${\rm gcd}(p, m_i)\neq 1$, then $f_i\in \varphi_i(\mathscr{C})$ can be selected to be of the form $f_i=b_i(x)(g_i(x)+up_i(x))+uc_i(x)a_i(x)$, where $g_i(x)\mid (x^{m_i}-1)$ over $\mathbb{F}_q$ and $a_i(x)\mid p_i(x)(\frac{x^{m_i}-1}{g_i(x)})$ with ${\rm deg}a_i(x)> {\rm deg}p_i(x)$.

\vskip 3mm \noindent {\bf Theorem 4.1 } \emph{Let $\mathscr{C}$ be a $1$-generator GQC code of block length $(m_1, m_2, \ldots, m_\ell)$ and length $\sum_{i=1}^\ell m_i$ over $R$. Let $G=(f_1g_1+uq_1, f_2g_2+uq_2, \ldots, f_\ell g_\ell+uq_\ell)$ be the generator of $\mathscr{C}$ and there exits $i\in \{ 1,2,\ldots,\ell\}$ such that $f_ig_i+uq_i$ is not a zero divisor of $R_i$. Suppose $h_i=\frac{x^{m_i}-1}{{\rm gcd}(f_ig_i, x^{m_i}-1)}$, $h={\rm lcm}\{h_1, h_2, \ldots, h_\ell\}$ with ${\rm deg}(h)=r$ and $v_i=\frac{x^{m_i}-1}{{\rm gcd}(hq_i, x^{m_i}-1)}$, $v={\rm lcm}(v_1, v_2, \ldots, v_\ell)$ with ${\rm deg}v=t$. Let $B=\{uhq_1, uhq_2, \ldots, uhq_\ell\}$. Then $\mathscr{C}$ has the minimal generating set $S_1\cup S_2$, where $S_1=\{ G, xG, \ldots, x^{r-1}G\}$, $S_2=\{B, xB, \ldots, x^{t-1}B\}$. Moreover, the codewords' number of $\mathscr{C}$ is $\mid \mathscr{C}\mid=q^{2r}q^t$.}

\vskip 3mm \noindent\emph{ Proof} Let $c(x)$ be a codeword of $\mathscr{C}$. Then $c(x)=f(x)G$, where $f(x)\in R[x]$. Since $h(x)$ is the regular polynomial, there are polynomials $Q_1(x), T_1(x)\in R[x]$ such that $f(x)=h(x)Q_1(x)+T_1(x)$ with $T_1(x)=0$ or ${\rm deg}T_1(x)\leq r-1$. Hence $c(x)=f(x)G=(h(x)Q_1(x)+T_1(x))(f_1g_1+uq_1, f_2g_2+uq_2, \ldots, f_\ell g_\ell+uq_\ell)=Q_1(x)B+T_1(x)G$. Note that $T_1(x)G\in {\rm Span}(S_1)$. Since $v(x)$ is also the regular polynomial, there are polynomials $Q_2(x), T_2(x)\in R[x]$ such that $Q_1(x)=Q_2(x)v(x)+T_2(x)$ with $T_2(x)=0$ or ${\rm deg}T_2(x)\leq t-1$. Therefore $Q_1(x)B=(Q_2(x)v(x)+T_2(x))B=T_2(x)B$. Clearly, $T_2(x)B\in {\rm Span}(S_2)$. Hence, $c(x)\in {\rm Span}(S_1)\cup {\rm Span}(S_2)$. Thus $\mathscr{C}$ can be generated by the set $S_1\cup S_2$.

\vskip 3mm Next we will prove ${\rm Span}(S_1)\cap {\rm Span}(S_2)=\{0\}$. Suppose $e(x)\in {\rm Span}(S_1)\cap {\rm Span}(S_2)$, where $e(x)=(e_1(x), e_2(x), \ldots, e_\ell(x))$, $e_i(x)\in R_i$ for each $i=1,2,\ldots,\ell$. Since $e(x)\in {\rm Span}(S_1)$, $e_i(x)=(f_ig_i+uq_i)M_i^{(1)}(x)$, where $M_i^{(1)}(x)=\alpha_0+\alpha_1x+\cdots+\alpha_{r-1}x^{r-1} \in R[x]$. On the other hand, $e_i(x)\in {\rm Span}(S_2)$, this implies that $e_i(x)=uhq_iM_i^{(2)}$, where $M_i^{(2)}=\beta_0+\beta_1x+\cdots+\beta_{t-1}x^{t-1}\in \mathbb{F}_q[x]$. Since $u^2=0$ in $R$, $ue_i(x)=u(f_ig_i+uq_i)M_i^{(1)}=0$. This implies that $\alpha_j=0$ or $\alpha_j=u$ for all $j=0,1,\ldots,r-1$. Note that $(f_ig_i+uq_i)M_i^{(1)}=uhq_iM_i^{(2)}$. Therefore $(f_ig_i+uq_i)(M_i^{(1)}-hM_i^{(2)})=0$. Since $f_ig_i+uq_i$ is not a zero divisor of $R_i$, we have $M_i^{(1)}-hM_i^{(2)}=0$ which deduces $\alpha_j=0$ and $\beta_i=0$ for all $j=0,1,\ldots,r-1$ and $i=0,1,\ldots,t-1$.         \hfill $\Box$

\vskip 3mm \noindent {\bf Corollary 4.2 } \emph{Let $\mathscr{C}$ be the GQC code in Theorem 4.1. If for each $i=1,2,\ldots,\ell$, the polynomial $f_ig_i+uq_i$ is a factor of $x^{m_i}-1$, then $\mathscr{C}$ is the free GQC code with ${\rm rank}(\mathscr{C})=r$ and $\mid \mathscr{C}\mid=q^{2r}$.}

\vskip 3mm \noindent\emph{ Proof} Since $f_ig_i+uq_i\mid (x^{m_i}-1)$ over $R$, it follows that $f_ig_i\mid (x^{m_i}-1)$ over $\mathbb{F}_q$. Let $h_i=\frac{x^{m_i}-1}{f_ig_i}$ over $\mathbb{F}_q$. Then there exists a polynomial $w_i\in \mathbb{F}_q[x]$ such that $\frac{x^{m_i}-1}{f_ig_i+uq_i}=h_i+uw_i$ over $R$. So there is a polynomial $s\in \mathbb{F}_q[x]$ such that $h+us={\rm lcm}\{h_1+uw_1, h_2+uw_2, \ldots, h_\ell+uw_\ell\}$, where $h={\rm lcm}\{ h_1, h_2, \ldots, h_\ell\}$. Thus $(h+us)(f_ig_i+uq_i)=0$ in $R_i$, which implies $u(f_ig_is+hq_i)=0$ in $R_i$. It means $hq_i=-f_ig_is$, which deduces $S_2\in {\rm Span}(S_1)$. Therefore $\mathscr{C}$ is free with ${\rm rank}(\mathscr{C})=r$ and the numbers' of codewords of $\mathscr{C}$ is $q^{2r}$.   \hfill $\Box$

\vskip 3mm In the following, we give a lower bound on the minimum Lee (Hamming) distance of the free $1$-generator GQC code over $R$.

\vskip 3mm \noindent {\bf Theorem 4.3}\emph{ Let $\mathscr{C}$ be a free $1$-generator GQC code as in Corollary 4.2. Suppose $h_i=\frac{x^{m_i}-1}{f_ig_i+uq_i}$, $i=1,2,\ldots,\ell$, and $h={\rm lcm}\{h_1,h_2,\ldots,h_\ell\}$. Then \\
$(i)$~$d_{\rm min}(\mathscr{C})\geq \sum_{{i}\not \in K}d_i$, where $K\subseteq \{1,2,\ldots,\ell\}$ is a set of maximum size for which ${\rm lcm}\{ h_i, i\in K\}\neq h$ and $d_i=d_{\rm min}(\varphi_i(\mathscr{C}))$;\\
$(ii)$~If $h_1=h_2=\cdots =h_\ell$, then $d_{\rm min}(\mathscr{C})\geq\sum_{i=1}^\ell d_i$.}

\vskip 3mm \noindent\emph{ Proof} Let $c(x)\in \mathscr{C}$ be a nonzero codeword. Then there exists a polynomial $f(x)\in R[x]$ such that $c(x)=f(x)G$. Since for each $i=1,2,\ldots,\ell$, $f_ig_i+uq_i\mid (x^{m_i}-1)$, the $i$-th component is zero if and only if $x^{m_i}-1\mid f(x)(f_ig_i+uq_i)$, that is, if and only if $h_i\mid f(x)$. Therefore $c(x)=0$ if and only if $h\mid f(x)$. So $c(x)\neq 0$ if and only if $h\nmid f(x)$. This implies that $c(x)\neq 0$ has the most number of zero blocks whenever $h\neq {\rm lcm}_{i\in K}h_i$, ${\rm lcm}_{i\in K}h_i\mid f(x)$, and $K$ is a maximal subset of $\{1,2,\ldots,\ell\}$ having this property. Thus, $d_{\rm min}(\mathscr{C})\geq \sum_{i\notin K}d_i$, where $d_i=d_{\rm min}(\varphi_i(\mathscr{C}))$. Clearly, $K=\emptyset$ if and only if $h_1=h_2=\cdots =h_\ell$. Therefore, from the discussion above, we have if $h_1=h_2=\cdots =h_\ell$, then $d_{\rm min}(\mathscr{C})\geq\sum_{i=1}^\ell d_i$.                                                                                 \hfill $\Box$

\vskip 3mm
By Corollary 4.2 and Theorem 4.3, we can get the following results immediately. Here we omit the proof.

\vskip 3mm \noindent {\bf Corollary 4.4}\emph{ Let $\mathscr{C}$ be a $1$-generator GQC code generated by $G=((f_1g_1+uq_1)k_1, (f_2g_2+uq_2)k_2, \ldots, (f_\ell g_\ell+uq_\ell)k_\ell)$, where for each $i=1,2,\ldots,\ell$, $f_ig_i+uq_i\mid (x^{m_i}-1)$. Suppose ${\rm deg}(f_ig_i+uq_i)={\rm deg }(f_ig_i)$, $h_i=\frac{x^{m_i}-1}{f_ig_i+uq_i}$, ${\rm gcd}(h_i, k_i)=1$, $i=1,2,\ldots,\ell$, and $h={\rm lcm}\{h_1,h_2,\ldots,h_\ell\}$. Then \\
$(i)$~$\mathscr{C}$ is free with rank degree of $h$ and $\mid \mathscr{C}\mid=q^{2{\rm deg}(h)}$; \\
$(ii)$~$d_{\rm min}(\mathscr{C})\geq \sum_{{i}\not \in K}d_i$, where $K\subseteq \{1,2,\ldots,\ell\}$ is a set of maximum size for which ${\rm lcm}\{ h_i,i\in K\}\neq h$ and $d_i=d_{\rm min}(\varphi_i(\mathscr{C}))$;\\
$(iii)$~If $h_1=h_2=\cdots =h_\ell$, then $d_{\rm min}(\mathscr{C})\geq\sum_{i=1}^\ell d_i$.}   \hfill $\Box$

\vskip 3mm \noindent {\bf Corollary 4.5} \emph{Let $\mathscr{C}_1$ be a free $1$-generator GQC code as in Theorem 4.1 (ii) and $\mathscr{C}_2=(fg+uq)$ be a free cyclic code with length $n$ over $R$ and $h=\frac{x^n-1}{fg+uq}$. Let $\mathscr{C}$ be a code obtained by concatenating of $\mathscr{C}_1$ and $\mathscr{C}_2$. Then \\
$(i)$~If ${\rm gcd}(h, h_i)=1$, then $\mathscr{C}$ is a free $1$-generator GQC code of length $\sum_{i=1}^\ell m_i+n$ with rank ${\rm deg}(h_ih)$ and $d_{\rm min}(\mathscr{C})\geq {\rm min}\{d_{\rm min}(\mathscr{C}_1), d_{\rm min}(\mathscr{C}_2)\}$.\\
$(ii)$~If $h\mid h_i$, then $\mathscr{C}$ is a free $1$-generator GQC code of  length $\sum_{i=1}^\ell m_i+n$ with rank ${\rm deg}(h_i)$ and $d_{\rm min}(\mathscr{C})\geq d_{\rm min}(\mathscr{C}_1)$.}          \hfill $\Box$

\vskip 3mm In the rest of this section, let us present some applications of these theorems. In Example 4.6, some good (optimal or suboptimal) linear codes over finite fields are obtained by the GQC codes over $R$.

\vskip 3mm \noindent {\bf Example 4.6} Considering the Gray map $\phi:~(\mathbb{F}_q+u\mathbb{F}_q)^n\rightarrow \mathbb{F}_q^{2n}$, defined by $\phi (x)=(\omega (x_1), \omega(x_2), \ldots, \omega(x_n))$, where $\omega(a+bu)=(b, a+b)$ for $a+bu \in \mathbb{F}_q+u\mathbb{F}_q$, $a,b\in \mathbb{F}_q$. In fact, this Gray map leads to the $(\textbf{u}|\textbf{u}+\textbf{v})$ well known construction of linear codes over finite fields
\vskip 1mm  (1) Let $\mathscr{C}$ be a $1$-generator GQC code of block length $(2,4)$ and length $6$ over $\mathbb{F}_2+u\mathbb{F}_2$. Let $G=(x+1+u, x^3+x^2+x+1+u)$ be the generator of $\mathscr{C}$. Then $h=x+1$ and $v=x^3+x^2+x+1$. Therefore, by Theorem 4.1, $\mathscr{C}$ has the minimal generating set $\{(x+1+u, x^3+x^2+x+1+u)\}\cup \{(u(1+x), u(1+x)), (u(1+x), u(1+x^2)), (u(1+x), u(x^2+x^3))\}$ and $|\mathscr{C}|=2^{2+3}=2^5$. From the Gray map $\phi$, we get $\phi(\mathscr{C})$ is the $[12,5,4]$ linear code over $\mathbb{F}_2$, which is optimal. The Hamming weight enumerator of $\phi(\mathscr{C})$ is $$W(\phi (\mathscr{C}))=x^{12}+7x^8y^4+16x^6y^6+7x^4y^8+y^{12}.$$

\vskip 1mm  (2) Let $\mathscr{C}$ be a $1$-generator GQC code of block length $(2,2)$ and length $4$ over $\mathbb{F}_2+u\mathbb{F}_2$. Let $G=(1+u, ux+u+1)$. Then $\phi (\mathscr{C})$ is the $[8,4,4]$ linear code over $\mathbb{F}_2$, which is optimal. The Hamming weight enumerator of $\phi (\mathscr{C})$ is $$W(\phi(\mathscr{C}))=x^8+14x^4y^4+y^8.$$

\vskip 1mm  (3)  Let $\mathscr{C}$ be a $1$-generator GQC code of block length $(2,3)$ and length $5$ over $\mathbb{F}_2+u\mathbb{F}_2$. Let $G=(1+u, x^2+1+u)$. Then $\phi (\mathscr{C})$ is the $[10,8,2]$ linear code over $\mathbb{F}_2$, which is optimal. The Hamming weight enumerator of $\phi (\mathscr{C})$  is $$W(\phi (\mathscr{C}))=x^{10}+12x^8y^2+36x^7y^3+46x^6y^4+60x^4y^6+28x^3y^7+9x^2y^8+4xy^9.$$

\vskip 1mm  (4) Let $\mathscr{C}$ be a $1$-generator GQC code of block length $(2,4)$ and length $6$ over $\mathbb{F}_2+u\mathbb{F}_2$. Let $G=(1+u, ux^2+(1+u)x+1+u)$. Then $\phi (\mathscr{C})$ is the $[12,6,4]$ linear code over $\mathbb{F}_2$, which is optimal. The Hamming weight enumerator of $\phi (\mathscr{C})$ is $$W(\phi(\mathscr{C}))=x^{12}+6x^8y^4+24x^7y^5+16x^6y^6+9x^4y^8+8x^3y^9.$$

\vskip 1mm  (5) Let $\mathscr{C}$ be a $1$-generator GQC code of block length $(2,2)$ and length $4$ over $\mathbb{F}_3+u\mathbb{F}_3$. Let $G=(1+u, ux+1+u)$. Then $\phi (\mathscr{C})$ is the $[8,4,4]$ linear code over $\mathbb{F}_3$, which is optimal. The Hamming weight enumerator of $\phi (\mathscr{C})$ is $$W(\phi(\mathscr{C}))=x^7y+24x^4y^4+16x^3y^5+32x^2y^6+8y^8.$$

\vskip 1mm  (6) Let $\mathscr{C}$ be a $1$-generator GQC code of block length $(2,3)$ and length $5$ over $\mathbb{F}_3+u\mathbb{F}_3$. Let $G=(2ux+1+u, 2ux^2+2ux+1+u)$. Then $\phi (\mathscr{C})$ is the $[10,8,2]$ linear code over $\mathbb{F}_3$, which is optimal. The Hamming weight enumerator of $\phi (\mathscr{C})$ is $$W(\phi(\mathscr{C}))=x^{10}+40x^8y^2+40x^7y^3+460x^6y^4+820x^5y^5 $$ $$+1600x^4y^6+1600x^3y^7+1300x^2y^8+600xy^9+100y^{10}.$$

\vskip 1mm  (7) Let $\mathscr{C}$ be a $1$-generator GQC code of block length $(3,3)$ and length $6$ over $\mathbb{F}_3+u\mathbb{F}_3$. Let $G=(x^2+x+1+u, x^2+(1+u)x+1+u)$. Then $\phi (\mathscr{C})$ is the $[12,4,6]$ linear code over $\mathbb{F}_3$, which is optimal. The Hamming weight enumerator of $\phi (\mathscr{C})$ is $$W(\phi(\mathscr{C}))=x^{11}y+10x^6y^6+12x^5y^7+36x^4y^8+12x^3y^9+6x^2y^{10}+4y^{12}.$$

\vskip 1mm  (8) Let $\mathscr{C}$ be a $1$-generator GQC code of block length $(2,3)$ and length $5$ over $\mathbb{F}_5+u\mathbb{F}_5$. Let $G=(1+u, ux+1+u)$. Then $\phi (\mathscr{C})$ is the $[10,8,2]$ linear code over $\mathbb{F}_5$, which is optimal. The Hamming weight enumerator of $\phi (\mathscr{C})$ is $$ W(\phi(\mathscr{C}))=x^{10}+56x^8y^2+252x^7y^3+2208x^6y^4+10072x^5y^5$$ $$+34820x^4y^6+78764x^3y^7 +117168x^2y^8+105512xy^9 +41772y^{10}.$$

\vskip 3mm \noindent {\bf Example 4.7} Define the Lee weight of the elements $0, 1, u, 1+u$ of $R=\mathbb{F}_2+u\mathbb{F}_2$ as $0, 1, 2, 1$, respectively. Moreover, the Lee weight of an $n$-tuple in $R^n$ is the sum of the Lee weights of its components. The Gray map $\phi$ sends the elements $0,1,u,1+u$ of $R$ to $(0,0), (0,1), (1,1), (1,0)$ over $\mathbb{F}_2$, respectively. It is easy to verify that $\phi$ is a linear isometry form $R^n$ (Lee distance) to $\mathbb{F}_2^{2n}$ (Hamming distance).
\vskip 1mm  (1) Let $\mathscr{C}$ be a free 1-generator GQC code of block length (3,4) and length $7$ generated by $(1+x+x^2,1+x^2)$ over $\mathbb{F}_2+u\mathbb{F}_2$. Then we have $h_1=1+x, h_2=(1+x)^2,h=(1+x)^2,d_{min}(\varphi_1(\mathscr{C}))=1,d_{min}(\varphi_2(\mathscr{C}))=2$. From Theorem 4.3, $d_{min}(\mathscr{C}) \ge 2$. In fact,  the minimum Lee distance of $\mathscr{C}$ is $d_{min}(\mathscr{C}) = 2$ actually.

\vskip 1mm (2) Let $\mathscr{C}$ be a free 1-generator GQC code of block length (4,6) and length $10$ generated by $(1+x^2,(1+x+x^2)^2)$ over $\mathbb{F}_2+u\mathbb{F}_2$. Then we have $h_1=(1+x)^2, h_2=(1+x)^2,h=(1+x)^2,d_{min}(\varphi_1(\mathscr{C}))=2,d_{min}(\varphi_2(\mathscr{C}))=3$. From Theorem 4.3, $d_{min}(\mathscr{C}) \ge 5$. In fact,  the minimum Lee distance of $\mathscr{C}$ is $d_{min}(\mathscr{C}) = 5$ actually.

\section{QC codes}
\vskip 3mm \noindent

Quasi-cyclic (QC) codes form a special class of GQC codes with $m_1=m_2=\cdots =m_\ell=m$, i.e., the QC code $\mathscr{C}$ of length $m\ell$ with index $\ell$ is an $R[x]$-submodule of $(R[x]/(x^m-1))^\ell$. Therefore Theorems 3.1, 3.3, 4.1, 4.3 for GQC codes in this paper can be applied to QC codes naturally. But in this section we use another point of view presented in \cite{Lally} to study QC codes over $R$. The duals of QC codes are also discussed briefly.

\vskip 3mm For convenience, we write an element $v\in R^{m\ell}$ as an $\ell$-tuple $v=(v_0, v_1, \ldots, v_{m-1})$, where $v_i=(v_{i,0}, v_{i,1}, \ldots, v_{i,\ell-1}) \in R^\ell$. Let the map $T^\ell$ on $R^{m\ell}$ be defined as follows $$T^\ell(v_0, v_1, \ldots, v_{m-1})=(v_{m-1}, v_0, \ldots, v_{m-2}).$$ Define a one-to-one correspondence $$\eta: R^{m\ell}\rightarrow (R[x]/(x^m-1))^\ell,$$ $$(v_{0,0}, v_{0,1}, \ldots, v_{0,\ell-1},v_{1,0}, v_{1,1}, \ldots, v_{1,\ell-1}, \ldots, v_{m-1,0}, v_{m-1,1}, \ldots, v_{m-1,\ell-1})$$$$\mapsto  v(x)=(v_0(x), v_1(x), \ldots,  v_{\ell-1}(x)),$$ where $v_j(x)=\sum_{i=0}^{m-1}v_{i,j}x^i$ for $j=0,1,\ldots,\ell-1$. Then the QC code $\mathscr{C}$ of length $m\ell$ with index $\ell$ defined above is equivalent to a linear code of length $m\ell$ over $R$, which is invariant under the map $T^\ell$. This definition of the QC code is known as conventional row circulant.

\vskip 3mm Let $v=(v_{0,0}, v_{0,1},\ldots,v_{0,\ell-1}, v_{1,0}, v_{1,1}, \ldots,v_{1,\ell-1},\ldots, v_{m-1,0}, v_{m-1,1}, \ldots, \\v_{m-1,\ell-1} )\in R^{m\ell}$ and $\widetilde{R}$ be the $\ell$-th Galois extension of $R$. Define an isomorphism between $R^{m\ell}$ and $\widetilde{{\mathcal R}}^m$ by associating with each $\ell$-tuple $(v_{i,0}, v_{i,1}, \ldots,v_{i,\ell-1})$, $i=0,1,\ldots,m-1$, and the element $v_i\in \widetilde{{\mathcal R}}$ represented as $v_i=v_{i,0}+v_{i,1}\xi+\cdots+v_{i,\ell-1}\xi^{l-1}$ where the set $\{1, \xi, \xi^2, \ldots, \xi^{\ell-1}\}$ forms an $R$-basis of $\widetilde{{\mathcal R}}$. Then every element in $R^{m\ell}$ is one-to-one correspondence with an element of $\widetilde{{\mathcal R}}^m$. The operator $T^\ell$ for some element $$(v_{0,0}, v_{0,1},\ldots,v_{0,\ell-1}, v_{1,0},v_{1,1}, \ldots, v_{1,\ell-1},\ldots, v_{m-1,0}, v_{m-1,1}, \ldots,v_{m-1,\ell-1} )\in R^{m\ell}$$ corresponds to the element $(v_{m-1}, v_0,\ldots, v_{m-2})\in \widetilde{{\mathcal R}}^m$. Indicating the block positions with increasing powers of $x$, the vector $v\in R^{m\ell}$ can be associated with the polynomial $v_0+v_1x+\cdots+v_{m-1}x^{m-1}\in \widetilde{{\mathcal R}}[x]$. Clearly, there ia an $R[x]/(x^m-1)$-module isomorphism between $R^{m\ell}$ and $\widetilde{{\mathcal R}}[x]/( x^m-1)$, which is defined as $\phi(v)=v_0+v_1x+\cdots+v_{m-1}x^{m-1}$. In this setting, multiplication by $x$ of any element of $\widetilde{{\mathcal R}}[x]/( x^m-1)$ is equivalent to applying $T^\ell$ to the element of $R^{m\ell}$. It follows that there is a one-to-one correspondence between the $R[x]/( x^m-1)$-submodule of $\widetilde{{\mathcal R}}[x]/( x^m-1)$ and the QC code of length $m\ell$ with index $\ell$ over $R$. In addition, let $\mathscr{C}$ be a QC code of length $m\ell$ with index $\ell$ over $R$. It can also be regarded as an $R$-submodule of $\widetilde{{\mathcal R}}[x]/( x^m-1)$ because of the equivalence of $R^{m\ell}$ and $\widetilde{{\mathcal R}}[x]/( x^m-1)$.

\vskip 3mm \par Let $\mathscr{C}$ be a QC code of length $m\ell$ with index $\ell$ over $R$ and generated by elements $v_1(x), v_2(x),\ldots, v_\rho(x)\in {\widetilde{\mathcal R}}[x]/( x^m-1)$ as an $R[x]/( x^m-1)$-submodule of ${\widetilde{\mathcal R}}[x]/( x^m-1)$. Then $\mathscr{C}=\{a_1(x)v_1(x)+a_2(x)v_2(x)+\cdots+a_\rho(x)v_\rho(x)|~a_i(x)\in R[x]/( x^m-1), i=1,2,\ldots,\rho\}$. As discussed above, $\mathscr{C}$ is also an $R$-submodule of ${\widetilde{\mathcal R}}[x]/(x^m-1)$. As an $R$-submodule of ${\widetilde{\mathcal R}}[x]/( x^m-1)$, $\mathscr{C}$ is generated by the set $$\{v_1(x), xv_1(x), \ldots, x^{m-1}v_1(x),  \ldots, v_\rho(x), xv_\rho(x), \ldots, x^{m-1}v_\rho(x)\}.$$

\vskip 3mm \par If $\mathscr{C}$ is generated by a single element $v(x)$ as an $R[x]/( x^m-1)$-submodule of ${\widetilde{\mathcal R}}[x]/( x^m-1)$, then $\mathscr{C}$ is called a \emph{$1$-generator} QC code. Let the preimage of $v(x)$ in $R^{m\ell}$ be $v$. Then for the $1$-generator QC code $\mathscr{C}$, we have $\mathscr{C}$ is generated by the set $\{v, T^\ell v, \ldots, T^{\ell(m-1)}v\}$. It is the conventional of row circulant definition of the $1$-generator QC code. In fact, let $v(x)=v_0+v_1x+\cdots+v_{m-1}x^{m-1}$ be a polynomial of ${\widetilde{\mathcal R}}[x]/( x^m-1)$, where $v_i=v_{i,0}+v_{i,1}\xi+\cdots+v_{i,\ell-1}\xi^{\ell-1}$, $i=0,1,\ldots,m-1$. Then $v(x)$ becomes an $\ell$-tuple of polynomials over $R$ with the fixed $R$-basis $\{1, \xi, \xi^2,\ldots,\xi^{\ell-1}\}$, where the degree of each component polynomial is at most $m-1$. Therefore, $v(x)$ becomes an element of $(R[x]/( x^m-1))^\ell$. So $\mathscr{C}$ is an $R[x]/( x^m-1)$-submodule of $(R[x]/( x^m-1))^\ell$. It is the conventional way of definition of the QC code.

\vskip 3mm \par Since $R[x]/( x^m-1)$ is a subring of $\widetilde{{\mathcal R}}[x]/( x^m-1)$ and $\mathscr{C}$ is an $R[x]/( x^m-1)$-submodule of $\widetilde{{\mathcal R}}[x]/( x^m-1)$, it follows that $\mathscr{C}$ is in particular a submodule of an $\widetilde{{\mathcal R}}[x]/( x^m-1)$-submodule of $\widetilde{{\mathcal R}}[x]/( x^m-1)$, i.e., the cyclic code $ \widetilde{\mathscr{C}}$ of length $m$ over $\widetilde{{\mathcal R}}$. Therefore, $d_{\rm min}(\mathscr{C})\geq d_{\rm min}(\widetilde{\mathscr{C}})$, where $d_{\rm min}(\mathscr{C})$ and $d_{\rm min}(\widetilde {\mathscr{C}})$ are the minimum distance of $\mathscr{C}$ and $\widetilde{\mathscr{C}}$ respectively. Lally \cite{Lally} has obtained another lower bound on the minimum Hamming distance of the QC code over finite fields. In the following, we generalized these results to QC codes over finite chain ring $R$.

\vskip 3mm \noindent {\bf Theorem 5.1} \emph{Let $\mathscr{C}$ be a $\rho$-generator QC code of length $m\ell$ with index $\ell$ over $R$ and generated by the set $\{v_i(x)=\widetilde{v}_{i,0}+\widetilde{v}_{i,1}x+\cdots +\widetilde{}v_{i,m-1}x^{m-1}, i=1,2,\ldots,\rho\}\subseteq \widetilde{{\mathcal R}}[x]/(x^m-1)$. Then $\mathscr{C}$ has a lower bound on the minimum Hamming distance given by $$d_{\rm min}(\mathscr{C})\geq d_{\rm min}(\widetilde{\mathscr{C}})d_{\rm min}(\mathscr{B}),$$ where $\widetilde{\mathscr{C}}$ is a cyclic code of length $m$ over $\widetilde{{\mathcal R}}$ generated by $v_1(x), v_2(x), \ldots,  \\v_\rho (x)$, and $\mathscr{B}$ is a linear code of length $\ell$ generated by $\{{\mathcal V}_{i,j}, i=1,2,\ldots, \rho, j=0,1,\ldots, m-1\}\subseteq R^\ell$ where each ${\mathcal V}_{i,j}$ is the vector corresponding to the coefficients $\widetilde{v}_{i,j} \in \widetilde{{\mathcal R}}$ with respect to an $R$-basis $\{1, \xi, \ldots, \xi^{\ell-1}\}$.}        \hfill $\Box$

\vskip 3mm  Since $\mathbb{F}_q$ is a subring of $R$ and the set $\{1,u\}$ forms a $\mathbb{F}_q$-basis of $R$, the discussion above can lead to a construction of $1$-generator QC codes with index $2$ over finite field $\mathbb{F}_q$ from cyclic codes over $R$.

\vskip 3mm Let $R=\mathbb{F}_q+u\mathbb{F}_q$, where $u^2=0$. Consider a cyclic code $\widetilde{\mathscr{C}}$ of length $m$ generated by a polynomial $v(x)$ over $R$. Let $\mathscr{C}$ be a linear code of length $2m$ spanned by the set $\{v(x), xv(x), \ldots, x^{m-1}v(x)\}$ over $\mathbb{F}_q$. Then $\mathscr{C}$ is a $1$-generator QC code of length $2m$ with index $2$. If $v(x)=v_0+v_1x+\cdots+v_{m-1}x^{m-1} \in R[x]/( x^m-1)$, then each $v_i$ is an $2$-tuple with respect to the fixed $\mathbb{F}_q$-basis $\{1,u\}$ of $R$. Now let the set $\{v_0, v_1, \ldots, v_{m-1}\}$ generate a linear code $\mathscr{B}$ of length $2$ over $\mathbb{F}_q$. Then from Theorem 5.1, we have

\vskip 3mm \noindent{\bf Corollary 5.2} \emph{Let $\mathscr{C}$ be a $1$-generator QC code of length $2m$ with index $2$ over finite field $\mathbb{F}_q$  generated by the set $\{ v(x),xv(x),\ldots,x^{m-1}v(x) \}$, where  $v(x)=v_0+v_1x+\cdots+v_{m-1}x^{m-1}\in R[x]/( x^m-1)$. Then
\vskip 1mm \noindent $(i)$ $\mathscr{C}$ has a lower bound on the minimum Hamming distance given by $$d_{\rm min}(\mathscr{C})\geq d_{\rm min}(\widetilde{\mathscr{C}})d_{\rm min}(\mathscr{B}),$$ where $\widetilde{\mathscr{C}}$ is a cyclic code of length $m$ over $R$ with generator polynomial $g(x)={\rm gcd}(v(x), x^m-1) \in R[x]/( x^m-1)$ and $\mathscr{B}$ is a linear code of length $2$ generated by $\{v_0, v_1, \ldots, v_{m-1}\}$ where each $v_i$ is an $2$-tuple with respect to a fixed $\mathbb{F}_q$-basis $\{1,u\}$ of $R$.}
\vskip 1mm \noindent $(ii)$ \emph{If the cyclic code $\widetilde{\mathscr{C}}$ in $(i)$ is free and the generator polynomial $g(x)$ has $\delta-1$ consecutive roots in some Galois extension ring of $R$, and if the set $\{v_0, v_1, \ldots, v_{m-1}\}$ generates a cyclic code $\mathscr{B}$ over finite field $\mathbb{F}_q$ of length $2$ with the minimum Hamming distance $\varepsilon$, then $$d_{\rm min}(\mathscr{C})\geq \delta\varepsilon.$$}                       \hfill $\Box$

\vskip 3mm \noindent {\bf Example 5.3} Let $R=\mathbb{F}_2+u\mathbb{F}_2$ and $R=\{0,1,u,\overline{u}=1+u\}$ where $u^2=0$. It is well known that $x^7-1=(x+1)(x^3+ux^2+x+\overline{u})(x^3+\overline{u}x^2+ux+\overline{u})$, where $x+1$, $x^3+ux^2+x+\overline{u}$ and $x^3+\overline{u}x^2+ux+\overline{u}$ are basic irreducible polynomials over $R$. Let $\widetilde{{\mathcal R}}=R[x]/(x^3+ux^2+x+\overline{u})$. Since $x^3+ux^2+x+\overline{u}$ is a basic primitive polynomial over $R$, the root $\xi$ of $x^3+ux^2+x+\overline{u}$ is a basic primitive element in $\widetilde{{\mathcal R}}$. Taking $v(x)=(x+1)(x^3+ux^2+x+\overline{u})=x^4+\overline{u}x^3+\overline{u}x^2+ux+\overline{u}$, then the cyclic code $\widetilde{\mathscr{C}}$ of length $7$ generated by $v(x)$ is free and the minimum Hamming distance of $\widetilde{\mathscr{C}}$ is at least $4$. The non-zero coefficients of $v(x)$ correspond to the elements $(1,1)$, $(0,1)$, $(1,1)$, $(1,1)$, $(1,0)$ with respect to the $\mathbb{F}_2$-basis $\{1,u\}$ of $R$ and they generate a cyclic code $\mathscr{B}$ of length $2$ with the minimum Hamming distance $1$ over $\mathbb{F}_2$. Therefore $\mathscr{C}$ is a $1$-generator QC code of length $14$, dimension $3$ and the minimum Hamming distance at least $4\times1=4$ over finite field $\mathbb{F}_2$. A generator matrix for $\mathscr{C}$ is given as follows
\begin{equation} \left(
  \begin{array}{ccccccc}
    11 & 01 & 11& 11 & 10& 00 & 00\\
    00 & 11 & 01& 11 & 11 & 10 & 00\\
    00 & 00 & 11& 01& 11 & 11& 10\\
  \end{array}
\right).
\end{equation}
In fact the minimum Hamming distance of $\mathscr{C}$ is $6$. Therefore $\mathscr{C}$ is a $1$-generator QC code with parameters $[14, 3, 6]$ over $\mathbb{F}_2$.

\vskip 3mm In the reset of this section, we discuss the duals of the QC codes over $R$. Define the Euclidean inner product of $u, v\in R^{m\ell}$ by $$u\cdot v=\sum_{i=0}^{m-1}\sum_{j=0}^{\ell-1}u_{i,j}v_{i,j}.$$ Let $\mathscr{C}$ be a QC code of length $m\ell$ with index $\ell$, $u\in \mathscr{C}$ and $v\in \mathscr{C}^\perp$. Since $\mathscr{C}$ is invariant under $T^\ell$ , we have $u\cdot T^\ell(v)=\sum_{i=0}^{m-1}u_i\cdot v_{i+m-1}=T^{(m-1)\ell}(u)\cdot v=0$, where $i+m-1$ is taken modulo $m$. Hence $T^\ell(v)\in \mathscr{C}^\perp$, which implies that the dual of the QC code $\mathscr{C}$ is also a QC code of the same index $\ell$.

\vskip 3mm  We define a conjugation map $^-$ on $R[x]$ such that $\overline{ax^i}=ax^{m-i}$, for $ax^i\in R[x]$. On $(R[x]/(x^m-1))^\ell$, let the Hermitian inner product of $a(x)=(a_0(x), a_1(x), \ldots, a_{\ell-1}(x))$ and $b(x)=(b_0(x), b_1(x), \ldots, b_{\ell-1}(x))\in (R[x]/(x^m-1))^\ell$ be $$\langle a(x), b(x)\rangle=\sum_{i=0}^{\ell-1}a(x)\cdot \overline{b_i(x)}.$$ By generalizing Proposition 3.2 in \cite{Ling 1}, we get

\vskip 3mm \noindent {\bf Proposition 5.4} \emph{Let $u, v \in R^{m\ell}$ and $u(x)$ and $v(x)$ be their polynomial representations of $(R[x]/(x^m-1))^\ell$, respectively. Then $T^{\ell k}(u)\cdot v=0$ for all $0\leq k \leq m-1$ if and only if $\langle u(x), v(x)\rangle=0$}                             \hfill $\Box$

\vskip 3mm Let $\mathscr{C}$ be a QC code of length $m\ell$ with index $\ell$ over $R$. Then from Proposition 5.4, $$\mathscr{C}^\perp=\{v(x)\in (R[x]/(x^m-1))^\ell\mid \langle c(x), v(x)\rangle=0,~\forall c(x)\in \mathscr{C}\}.$$ Furthermore, if $\mathscr{C}$ is a QC code of length $m\ell$ with index $\ell$ over $R$, then from Corollary 3.2, $$\mathscr{C}^\perp=\bigoplus_{i=1}^s\mathscr{C}_i^\perp.$$

\vskip 3mm In \cite{Ling 3}, some results for $\rho$-generator QC codes and their duals over finite fields are given. These results can also be generalized to  $\rho$-generator QC codes over $R$.

\vskip 3mm \noindent {\bf Theorem 5.5}\emph{ Let $\mathscr{C}$ be a $\rho$-generator QC code of length $m\ell$ with index $\ell$ over $R$. Let $\mathscr{C}=\bigoplus_{i=1}^s\mathscr{C}_i$, where each $\mathscr{C}_i$, $i=1,2,\ldots,s$, is free and with rank $k_i$. Then \\
$(i)$~$\mathscr{C}$ is a ${\mathcal K}$-generator QC code and $\mathscr{C}^\perp$ is a $(\ell-{\mathcal K}')$-generator QC code, where ${\mathcal K}={\rm max}_{1\leq i \leq s}k_i$ and ${\mathcal K}'={\rm min}_{1\leq i\leq s}k_i$. \\
$(ii)$~Let $\ell\geq 2$. If $\mathscr{C}^\perp$ is also a $\rho$-generator QC code, then ${\rm min}_{1\leq i\leq s}k_i=\ell-\rho$ and $\ell\leq 2\rho$. \\
$(iii)$~If $\mathscr{C}$ is a self-dual $\rho$-generator QC code, then $\ell$ is even and $\ell\leq 2\rho$.  } \hfill $\Box$

\vskip 3mm For $1$-generator QC code of length $m\ell$ with index $\ell$ with canonical decomposition $\mathscr{C}=\bigoplus_{i=1}^s\mathscr{C}_i$, if each $\mathscr{C}_i$ is free, then $\mathscr{C}^\perp$ is also $1$-generator QC code if and only if $\ell=2$ and ${\rm rank}(\mathscr{C}_i)=1$ for each $i=1,2,\ldots,s$.

\section{Conclusion}
\vskip 6mm \noindent
Structural properties of the GQC code of arbitrary length are considered over the finite chain ring $\mathbb{F}_q+u\mathbb{F}_q$. Using the Chinese Remainder Theorem, we give some characteristics of the $\rho$-generator GQC code, which lead to a decomposition of the GQC code over $R$. For the $1$-generator GQC code, we give the minimal generating set and the free conditions. A lower bound on the minimum distance of the free $1$-generator GQC code is also given. Using the Gray map, some good (optimal or suboptimal) linear codes over finite field are obtained by this family, which implies that the GQC codes over $R$ are an interesting and useful class of linear codes. Finally, we study a special class of GQC codes called QC codes over $R$. We use another point of view to study the QC code, which leads to another distance bound for the QC code. The duals of QC codes are also discussed briefly.

\vskip 3mm In \cite{Cao 1}, Cao has given the explicit enumeration of all $1$-generator GQC codes and $1$-generator GQC codes with a fixed parity-check polynomial, respectively. These results are based on the fact that the ring $\mathbb{F}_q[x]/(g_k^{d_{ik}})$ is a finite chain ring. But, by Proposition 2.1, one can check that the ring $R[x]/(g_k^{d_{ik}})$ in this paper is not a finite chain ring anymore, since its local ideal $(u, g_k)$ is not principal. Therefore, what is the explicit enumeration of GQC codes of arbitrary lengths over $R$ is an interesting open problem for further consideration.
\vskip 3mm  It should be noted that most of the results in Section 3 and Section 4 in this paper can not be generalized to general finite chain rings. Therefore, the research of the structural properties of GQC codes of arbitrary lengths over general finite chain rings is also an interesting open problem for further consideration.

\vskip 3mm \noindent {\bf Acknowledgments}  \emph{The first author would like to thank Professor Cao for his comments on Theorem 3.1. This research is supported by the National Key Basic Research Program of China (Grant No. 2013CB834204), and the National Natural Science Foundation of China (Grant Nos. 61171082, 10990011 and 60872025).}

\bibliographystyle{model1a-num-names}

\bibliography{<your-bib-database>}



\end{document}